# Unraveling Saturable Absorption of Thulium-doped Fibres for New Level of Integrated Femtosecond Pulse Generation

Dennis C. Kirsch[1,*], Benoit Cadier[2], Thierry Robin[2], and Maria Chernysheva[1]

[1] *Leibniz Institute of Photonic Technology, Albert-Einstein-Str. 9, 07702 Jena, Germany*
[2] *iXblue Photonics, Rue Paul Sabatier 22300 Lannion, France*
\* *Corresponding author: dennis.kirsch@leibniz-ipht.de*

**The ultrashort pulse laser operation is crucially defined by engaging explicit modulators. Conventionally, they are laborious to incorporate into an all-fibre design or involve the design of sophisticated cavities. We exploit Thulium-doped fibre as the gain and simple, highly integrated saturable absorber with a modulation depth of 23% and a high thermal damage threshold. Hence, pulses as short as 350 fs carrying 1 nJ energy at a repetition rate of 45 MHz have been demonstrated. The laser average output power showed long-term stability with ∼1%$_{\text{RMSE}}$. The utilised concept of all-fibre saturable absorber provides technological grounds for a down-scaled, highly integrated laser architecture.**

Over the last decade, the ultrafast short-wave infrared (SWIR: 1.6 – 2.5 µm) laser market has been auspiciously growing with Thulium-doped fibre systems as key players. An advantageous low-loss atmospheric transmission, a deep biological tissue penetration as well as various absorption lines of gases and biomolecules, including $H_2O$, $CO_2$, $CH_4$, lipids or glucose, drive the demand on efficient light sources operating at this wavelength band. To unleash the full potential for expanding applications in medical examination and surgery, ranging, optical coherent tomography, as well as polymer processing or industrial laser cutting and welding, cost-effective, rugged, compact turn-key solutions have been sought after [1, 2]. However, it has become evident that a higher degree of fibre integration, particularly in criteria of footprint, scalability and tuneability, with low-cost performance, is imperative to solidify the rapid progress of the past decade.

In order to passively generate ultrafast pulsed radiation, phase-locking of longitudinal cavity modes is conventionally achieved through the implementation of certain components that exhibit saturable absorption behaviour. The most widespread saturable absorbers are quantum-well-based SESAMs (e.g.GaSb) [3, 4], but their performance at SWIR wavelength range lags behind their Near-IR counterparts. Moreover, SESAMs are cumbersome to integrate into the all-fibre design. Alternatively, research is being conducted on a wide variety of novel low dimensional materials, including but not limited to carbon nanotubes [5], transition metal dichalcogenides [6], MXene [7] or graphdyne [8]. Such low dimensional materials demonstrated superior fibre compatibility, despite having a low damage threshold, in general.

On the other hand, modulators based on the nonlinear optical Kerr effect, such as nonlinear polarisation evolution, nonlinear loop mirrors or nonlinear multimode interference, hold a faster recovery time, higher damage threshold and better integrability. Nonetheless, they suffer from the diminished nonlinearity of optical fibres at the SWIR band, leading to a high self-starting threshold and necessity to design complex cavity configurations [9].

While the former methods have proven their efficiency, there is an active search for novel cost-effective mode-locking techniques with high damage threshold, high integrability and wide operational range. This has brought out latent possibilities of rare-earth-doped fibres to act as mode-locking modulators. Indeed, rare-earth fibres exhibit saturable absorber behaviour when their absorption cross-section covers the emission cross-section of gain fibre. For example, Q-switched generation in Er-doped fibre laser was realised by implementing a section of Tm-doped fibre [10, 11], or Ho-doped fibre was used as a saturable absorber in Tm-doped fibre laser [12]. The slow recovery time of such saturable absorption makes it imperative to rely on a high degree of soliton shaping. Therefore, the pulse formation is assisted by the reciprocity of self-phase modulation (SPM) and anomalous group-delay dispersion [13].

Nevertheless, the phenomenon of self-mode-locking, with no apparent saturable absorber involved in the laser cavity, presents an even more attractive opportunity. In fibre oscillators, self-mode-locking was observed through dint of nonlinear coupling between cores in multi-core fibres [14], inter-mode beating [15], or self-absorption in the oscillator gain fibre.

Conventionally, the latter was considered as undesired instability in continuous-wave fibre lasers [16] and only later was studied as a mode-locking mechanism. The origin of the saturable absorption behaviour in rare-earth-doped fibres is associated with excited-state absorption [17–19] and upconversion interaction between ion pairs [20, 21]. Further, in the context of Thulium-doped fibres, *Jackson et. al.* have concluded that $^3H_4$, $^3H_4 \rightarrow {}^3H_6$, $^3F_4$ energy transfer upconversion process plays a key role in establishing the saturable absorption behaviour[22]. Due to the high energy difference of -1500 cm$^{-1}$ between the initial and the final upconversion energy states, the rate of this energy transfer process is weaker than in Erbium-doped fibres (∼600 cm$^{-1}$)[20]. Therefore, a higher degree of clustering is required for the Tm-doped silica system than for Er-doped silica to achieve ion separations that are short enough to provide a given level of saturable absorption. It is known that with the increase of doping concentration of rare-earth ions in silica fibres, ion-pair clusters are more likely to form. Therefore, to achieve an adequate value of saturable absorption modulation depth, a high concentration of $Tm^{3+}$ ions is required.

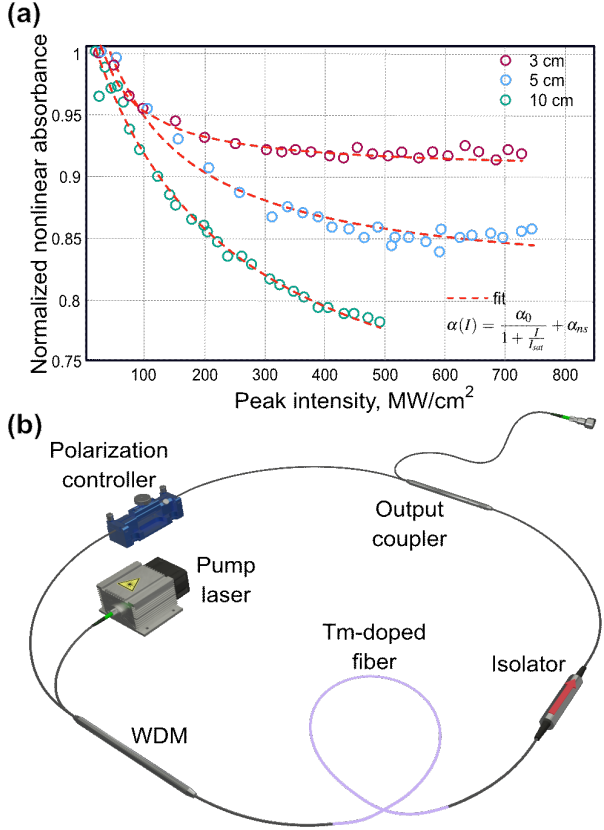

**Fig. 1.** a) Measurement of the normalised nonlinear absorption of a 5 cm-long Thulium-doped fibre via twin-detector approach – fit according to eq. **??**; b) schematic of the self-mode-locked Thulium-doped fiber laser – WDM: wavelength division multiplexer.

So far, there have been only few reports, where the gain fibre saturable absorber dynamic enabled the Q-switched generation [23] or the formation few-picosecond mode-locked pulses and their stabilisation against continuous wave disturbance [24, 25]. To ensure the required value of the SPM and anomalous group-delay dispersion for ultrashort pulse formation, the laser cavities had to be elongated to 50 [24] or even 95 m [25]. Nonetheless, it is worth highlighting the work by *Nyushkov et al.*, which took the Er-doped fibre laser cavity elongation to the extreme kilometre-long value and achieved the emission of 1.7 μJ, 35.1 kHz pulses via self-mode-locking [26].

In the current work, we explore a heavily-doped active fibre enriched with $Tm^{3+}$ ion clusters to reinforce its saturable absorption mechanism. Aided by this specialty fibre, the designed self-mode-locked ring fibre oscillator delivers near transform limited 350-fs soliton pulses at 45 MHz repetition rate with the energy of 1.1 nJ. The demonstrated approach has high potential only for development of integrated femtosecond fibre lasers operating at SWIR, but also other wavelength regions where the selection of efficient saturable absorbers is limited.

We studied a step-index non-polarisation-maintaining Thulium-doped fibre (from iXblue Photonics) with $1.3 \cdot 10^{26}\,m^{-3}$ ($0.3\%_{mol}$) doping concentration in an aluminosilica core with a 2.65-μm radius, leading to ~30 dB small-signal gain. The fibre featured 0.17 numerical aperture and an estimated group velocity dispersion of $-32\,ps^2\,km^{-1}$ at 1890 nm.

To measure the saturable absorption of the active fibre via the twin-detector approach, we used a self-build Thulium-doped fibre laser generating 500-fs pulses at 1890 nm. The laser output power, controlled via an external variable optical attenuator, was split into two arms and launched into an unpumped fibre under test, as well as a reference detector. Figure 1(a) shows the normalised intensity-dependent absorption of 3-, 5- and 10-cm long section of the used Tm-doped fibre with the approximation according to the two-level energy model (inset in Fig. 1a) [27], where $\alpha_0$ and $\alpha_{ns}$– are the modulation depth and non-saturable losses, correspondingly. $I$ – is the launched pulse intensity, and $I_{sat}$ – is the saturation intensity, which corresponds to the twice reduced absorption of the test sample. As seen in Fig. 1(a), the saturation intensity drops form 250 to 72 MW cm$^{-2}$) with the reduction of the active fibre length. It is also seen, that the 10-cm-long section of the fibre could not be fully saturated with the available laser power. The modulation depths of the tested 10- and 5-cm-long fibre sections are 34% to 19%, correspondingly. Overall, the modulation depth of the heavily doped fibre is comparable to the one of low dimensional material saturable absorbers [4, 6]. However, the saturation intensity is significantly higher, which can be explained by a high energy difference between the initial and final up-conversion energy states mentioned above. Further reduction of the Tm-doped fibre length leads to the decrease of modulation depth down to 9.5%.

The Tm-doped fibre laser arrangement is presented schematically in Fig. 1(b). The oscillator is built in a ring configuration with 0.5 m of the Thulium-doped fibre, which provided both active gain and saturable absorption to enable self-mode-locking. The pump energy is supplied through a 1550/2000 WDM by a continuous-wave Er-doped fibre laser with up to 1.04 W at 1549.6 nm, providing 0.04 nm linewidth and 1.7% power stability (HPFL-300, BKtel, France). The ring Thulium-doped fibre laser employs only a few standard non-polarisation-maintaining fiberised components: an isolator to ensure unidirectional generation, a squeezing polarisation controller to restore the polarisation state after each round trip, and a fused output coupler with 20% feedback. The length of passive single-mode fibre ports of the components is 4.1 m.

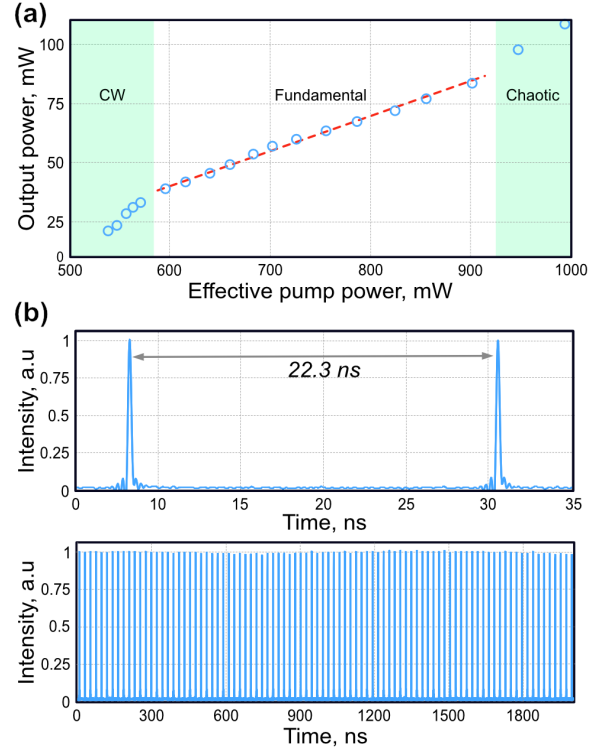

**Fig. 2.** a) Laser average power performance with linear fit of the slope efficiency to 14.5%; b) a pair of consequent pulses with spacing concordant to fundamental cavity round-trip time ; c) zoomed-out pulse train with 1%rms amplitude fluctuation.

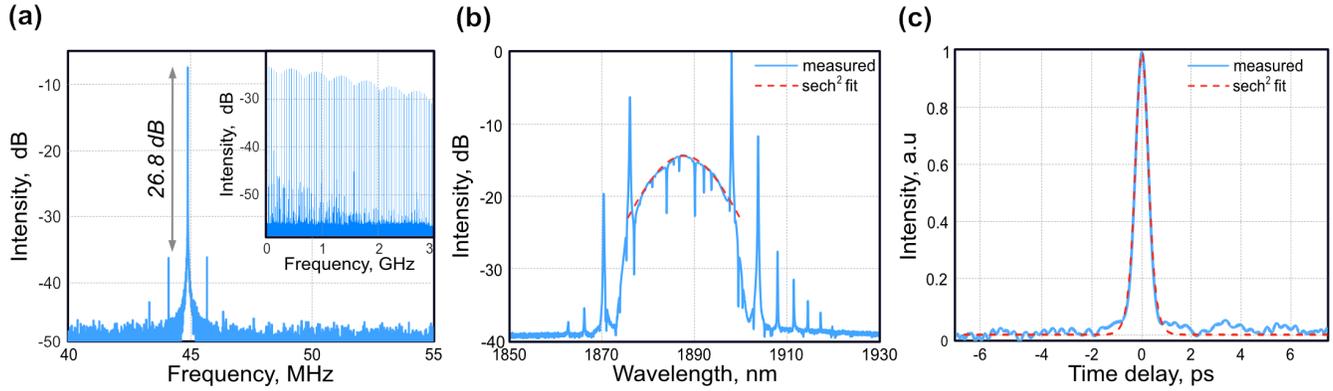

**Fig. 3.** Laser output characteristics at maximum 83 mW average output power: a) radio frequency spectrum of the fundamental note and wideband measurement in the inset; b) logarithmic optical spectrum; c) autocorrelation trace having 545 fs full width half maximum.

The laser continuous-wave generation threshold was 541 mW, while stable single-pulse mode-locked operation could be achieved at 596 mW (Fig. 2a). Such a relatively high mode-locking threshold can be attributed to the high saturation intensity (Fig. 1(a) of Tm-doped fibre when it acts as the mode-locking modulator. It is worth noticing that the laser required fine adjustment of the polarisation state at first, while after several switching on-off cycles, the mode-locking regime could self-start without polarisation tuning. With the fixed polarisation controllers, the single-pulse self-mode-locking could be maintained at up to 903 mW pump power. The maximum output power reached 83 mW, producing laser slope efficiency of 14.5%. In contrast to conventional ultrafast lasers, the efficiency of mode-locked regime is slower than of continuous wave generation. We attribute this to the nature of the saturable absorption mechanism in Tm-doped fibres associated with reabsorption of the laser signal. Further pump power increase led to the pulse break-up and chaotic generation regime.

The output pulse train, recorded by a 6-GHz oscilloscope (Tektronix, DPO 70604C) alongside a 12.5-GHz InGaAs photodetector (EOT, ET-5000F), is displayed over time in Fig. 2b. A broadband scan (lower plot) detects only minor amplitude fluctuations below 1%$_{rms}$, justifying a proper continuous-wave mode-locking. A zoomed-in plot displays a pulse sequence with a 22.3-ns repetition period, corresponding to the fundamental mode of the cavity length of ∼4.6 m. The pulse train features high-dynamic single amplitudes with no secondary pulsations.

The radio frequency spectrum in Fig. 3a justifies a pulse repetition rate of 44.8 MHz with an acceptable signal-to-noise ratio of 26.8 dB. The symmetrical sidebands with an offset of 0.78 MHz indicate a marginal variance in the repetition frequency, yet the spectrum base is rather flat. It should be noticed that at lower output powers, such modulation has not been observed. Therefore, we believe that the stability can be further improved with proper optimisation of the laser cavity setup.

As shown in Fig. 3b, the output optical spectrum is centered at 1888 nm. It features a classical soliton profile (red dashed curve), as expected for the laser cavity with fully anomalous dispersion. Its −3-dB spectral bandwidth measures 12 nm. Acquired during free-space propagation in the optical spectrum analyser (Yokogawa, AQ6375B), the spectrum is interlaced with water absorption lines. From the position of its characteristic Kelly sidebands, the net cavity dispersion amounts to -0.283 ps$^2$, which is in good agreement with the value derived from the fibre parameters and lengths. About 36% of the pulse power belongs to the sidebands, estimated by integrating the spectrum.

The intensity autocorrelation (APE, PulseCheck) presented in Fig. 3c features a lone symmetric peak without secondary signals or a substantial pedestal. The full width at half maximum of the trace is 545 fs. Due to the good agreement with the sech$^2$ profile, the pulse duration can be estimated as 350 fs. In terms of temporal and spectral width, the soliton is nearly transform-limited with a time-bandwidth product of 0.353. At the maximum pump power, the single pulse energy reaches 1.1 nJ, with the Kelly sidebands energy left out of considerations.

Further, the self-mode-locked Tm-doped fibre laser generation at maximum output performance was tested over 49 hours of continuous operation to assess its stability. Figure 4(a) illustrates the dynamics of the output pulse spectrum, recorded every 10 minutes. It justifies that no pulse breakup occurred, merely a negligible variance in the intensity of Kelly-sidebands. However, it is notable that the central wavelength blue-shifts by ∼1 nm (Fig. 4c) over the time of measurement. Importantly, the trend of the wavelength drift in Fig. 4(a) correlates with the arching of the output average power dynamics (Fig. 4b), showing an overall ∼1.0% standard deviation. The timing of laser performance alternation allows us to suggest that its origin lies in the temperature fluctuations of the laboratory.

To get a deeper insight into the performance of Tm-doped fibre as a saturable absorber, we have studied the self-mode-locked laser generation, gradually cutting back the length of the active fibre. Our findings demonstrate that the mode-locking dynamics are sensitive to the length of Tm-doped fibre. The ultrashort pulse generation could still be achieved with the length of Tm-doped fibre above 47.8 cm. However, Q-switch intensity modulation became more pronounced and affected the quality of the generation. The cut-back studies demonstrated a reduction of the spectral bandwidth of the generated solitons down to 0.95 nm, as shown in Fig. 4(d). It also revealed the raise of the mode-locking threshold from 610 mW effective pump power at the original length, over 690 mW to 810 mW. Furthermore, laser efficiency increased with the active fibre length reduction, as at the same gain distribution the losses due to signal reabsorption decreased. The fact that the reduction of the fibre length impacted only its unexcited part allows us to make a hypothesis that this fibre section serves as a saturable absorber. Unquestionably, shorter Tm-doped fibre possesses a lower saturable and non-saturable loss, which becomes insufficient for

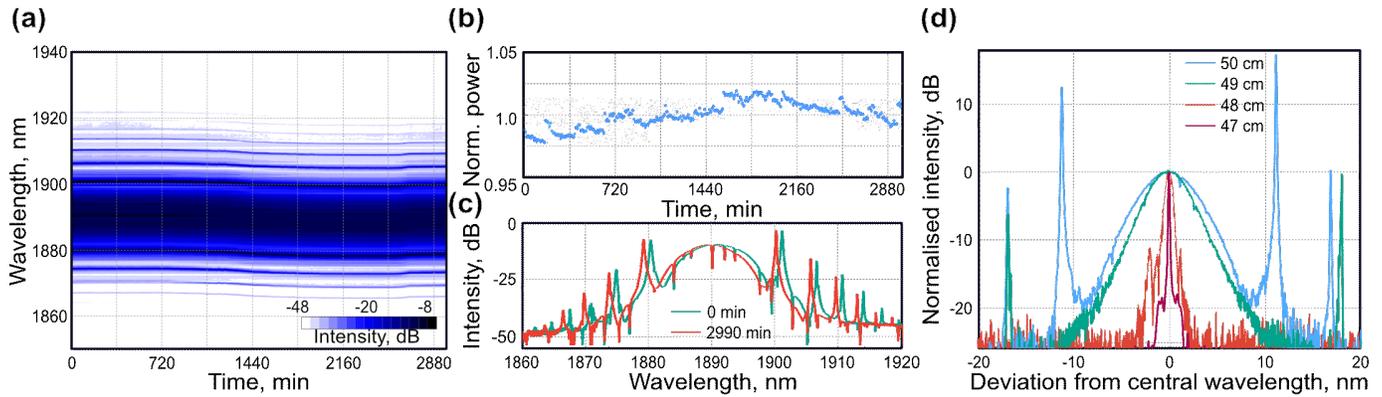

**Fig. 4.** (a-c) Long-term self-mode-locked laser generation stability at maximum 83 mW average output power: a) spectrum dynamics; b) average output power variation; c) wavelength drift during 49 h; (d) spectral evolution with the active fibre cut-back

stable ultrashort pulse formation in the laser cavity.

In conclusion, we have experimentally investigated the nonlinear absorption properties of Tm-highly-doped fibre. The high $1.3 \cdot 10^{26}$ m$^{-3}$ doping concentration introduces efficient upconversion interaction between ion pairs in the step-index Tm-doped fibre. This equips gain fibre with 23% modulation depth, yet high saturation intensity (95 MW·cm$^{-2}$). The demonstrated in the current work self-mode-locked laser setup enabled the generation of stable self-starting ultrashort pulses with the duration of 350 fs at 45 MHz repetition rate delivering over 1 nJ of the energy in the main pulse peak. To the best of our knowledge, this is the first demonstration of femtosecond pulse generation directly from the self mode-locked Thulium-doped fibre laser.

Therefore, the investigated approach simplifies the ultrafast fibre laser cavity design significantly since a single Tm-doped fibre element provides gain and enables saturable absorption. Moreover, decent soliton energy can be achieved directly from the cavity without the danger of thermal degradation of laser components. Additionally, this method may aid oscillator configurations insufficient for self-starting behaviour by hybrid mode-locking approach as a slow saturable absorber is more efficient for enabling self-starting than faster modulators. It should be acknowledged that the major technological complexity shifts to the design of appropriate active fibre compositions. The degradation of mode-locking operation during the active fibre cut-back studies made the decisive role of the optimal selection of fibre parameters even more evident. Detailed spectroscopic and pump-probe investigation of various Thulium-doped fibre compositions would be beneficial for in-depth evaluation of the impact of clustering and quenching processes on the relaxation and saturable absorption behaviour.

Possible strategies to further reduce the state lifetime include increasing the matrix phonon energy and non-linearity, e.g. by co-doping with Lanthana or Terbia. Other co-doping elements should be explored to obtain more substantial quenching of Tm$^{3+}$ ions.

To summarise, the reduction of the components required in the laser setup, and hence its price, push the R&D direction towards market-driven low-cost, compact SWIR photonics systems and, potentially, mid-IR sources.

**Disclosures.** The authors declare no conflicts of interest.

## REFERENCES

1. D. Kirsch, S. Chen, R. Sidharthan, Y. Chen, S. Yoo, and M. Chernysheva, J. Appl. Phys. **128**, 180906 (2020).
2. J. Ma, Z. Qin, G. Xie, L. Qian, and D. Tang, Appl. Phys. Rev. **6**, 021317 (2019).
3. J. Heidrich, M. Gaulke, B. O. Alaydin, M. Golling, A. Barh, and U. Keller, Opt. Express **29**, 6647 (2021).
4. M. Zhang, H. Chen, J. Yin, J. Wang, J. Wang, and P. Yan, Chin. Opt. Lett. **19**, 081405 (2021).
5. K. Watanabe, Y. Zhou, Y. Sakakibara, T. Saito, and N. Nishizawa, OSA Continuum **4**, 137 (2021).
6. X. Liu, Q. Gao, Y. Zheng, D. Mao, and J. Zhao, Nanophotonics **9**, 2215 (2020).
7. Z. Wang, H. Li, M. Luo, T. Chen, X. Xia, H. Chen, C. Ma, J. Guo, Z. He, Y. Song *et al.*, ACS Appl. Nano Mater. **3**, 3513 (2020).
8. J. Guo, Z. Wang, R. Shi, Y. Zhang, Z. He, L. Gao, R. Wang, Y. Shu, C. Ma, Y. Ge *et al.*, Adv. Opt. Mater. **8**, 2000067 (2020).
9. H. Haus, E. Ippen, and K. Tamura, IEEE J. Quantum Electron. **30**, 200 (1994).
10. H.-X. Tsao, C.-H. Chang, S.-T. Lin, J.-K. Sheu, and T.-Y. Tsai, Opt. & Laser Technol. **56**, 354 (2014).
11. C. XU, T. HUANG, Z. WU, L. HAN, J. ZHANG, and D. WANG, Appl. Opt. **60**, 6843 (2021).
12. J. Gene, S. K. Kim, and S. Do Lim, J. Light. Technol. **36**, 2183 (2018).
13. R. Paschotta and U. Keller, Appl. Phys. B **73**, 653 (2001).
14. H. G. Winful and D. T. Walton, Opt. letters **17**, 1688 (1992).
15. J. Zhang, D. Wu, H. Zhang, R. Zhao, R. Wang, and S. Dai, Opt. Fiber Technol. **58**, 102253 (2020).
16. F. Fontana, M. Begotti, E. Pessina, and L. Lugiato, Opt. communications **114**, 89 (1995).
17. P. Le Boudec, P. Francois, E. Delevaque, J.-F. Bayon, F. Sanchez, and G. Stephan, Opt. quantum electronics **25**, 501 (1993).
18. A. F. El-Sherif and T. A. King, Opt. communications **208**, 381 (2002).
19. Y. Tang and J. Xu, IEEE J. Quantum Electron. **47**, 165 (2011).
20. F. Sanchez, P. Le Boudec, P.-L. Francois, and G. Stephan, Phys. Rev. A **48**, 2220 (1993).
21. S. Colin, E. Contesse, P. Le Boudec, G. Stephan, and F. Sanchez, Opt. letters **21**, 1987 (1996).
22. S. D. Jackson and T. A. King, JOSA B **16**, 2178 (1999).
23. F. Z. Qamar and T. A. King*, J. Mod. Opt. **52**, 1053 (2005).
24. C. Liu, Z. Luo, Y. Huang, B. Qu, H. Cheng, Y. Wang, D. Wu, H. Xu, and Z. Cai, Appl. optics **53**, 892 (2014).
25. M. Wang, J. Zhao, Y. Li, and S. Ruan, Optik **154**, 485 (2018).
26. B. Nyushkov, V. Denisov, S. Kobtsev, V. Pivtsov, N. Kolyada, A. Ivanenko, and S. Turitsyn, Laser Phys. Lett. **7**, 661 (2010).
27. W. T. Silfvast, *Laser fundamentals* (Cambridge university press, 2004).